# Design and Demonstration of Ultra-Wide Bandgap AlGaN Tunnel Junctions

Yuewei Zhang,[1,a)] Sriram Krishnamoorthy,[1] Fatih Akyol,[1] Andrew A. Allerman,[2] Michael W. Moseley,[2] Andrew M. Armstrong,[2] and Siddharth Rajan[1,3,a)]

[1] Department of Electrical and Computer Engineering, The Ohio State University, Columbus, Ohio, 43210, USA

[2] Sandia National Laboratories, Albuquerque, New Mexico 87185, USA

[3] Department of Materials Science and Engineering, The Ohio State University, Columbus, Ohio, 43210, USA

**Abstract:** Ultra violet light emitting diodes (UV LEDs) face critical limitations in both the injection efficiency and light extraction efficiency due to the resistive and absorbing p-type contact layers. In this work, we investigate the design and application of polarization engineered tunnel junctions for ultra-wide bandgap AlGaN (Al mole fraction > 50%) materials towards highly efficient UV LEDs. We demonstrate that polarization-induced 3D charge is beneficial in reducing tunneling barriers especially for high composition AlGaN tunnel junctions. The design of graded tunnel junction structures could lead to low tunneling resistance below $10^{-3}$ Ω cm$^2$ and low voltage consumption below 1 V (at 1 kA/cm$^2$) for high composition AlGaN tunnel junctions. Experimental demonstration of 292 nm emission was achieved through non-equilibrium hole injection into wide bandgap materials with bandgap energy larger than 4.7 eV, and detailed modeling of tunnel junctions shows that they can be engineered to have low resistance, and can enable efficient emitters in the UV-C wavelength range.

a) Authors to whom correspondence should be addressed.
Electronic mail: zhang.3789@osu.edu, rajan@ece.osu.edu

The large range of direct bandgaps in the III-Nitride system makes these semiconductors uniquely suitable for achieving ultra violet emission over a wide wavelength range down to 210 nm. III-Nitride ultra-violet light emitting diodes (UV LEDs) have a large variety of applications including water purification and air disinfection, and they provide distinct advantages over traditional gas-based lamps.[1] However, while AlGaN-based UV LEDs with peak wavelength ranging between 210 nm and 365 nm have been demonstrated, the wall-plug efficiency (WPE) for UV LEDs below 365 nm is still significantly lower than that for visible GaN LEDs.[2-7] The underlying reason for this is the high acceptor ionization energy in AlGaN, which makes p-type contacts and layers highly resistive.

The conventional solutions to the p-type doping issue have involved using a combination of p-AlGaN supperlattice (SL) and p-GaN cap layers (Fig. 1(a)) as the hole injection layer.[2-7] However, the poor p-type conductivity of p-AlGaN leads to low injection efficiency and high operation voltage[8], while the p-GaN cap layer and metal contact absorb UV light, resulting in reduced light extraction[4,5].

We propose an approach to realize highly efficient hole injection based on interband tunneling. In this approach, a transparent n-type AlGaN layer is connected to the p-AlGaN layer using an interband tunnel junction (Fig. 1(b)). Non-equilibrium tunneling injection of holes overcomes limits imposed by thermal ionization of deep acceptors.[9,10] At the same time, it leads to efficient light extraction since it eliminates absorbing p-type top contact layers.[11]

Making efficient interband p-n tunnel junctions is challenging for ultra-wide bandgap AlGaN materials due to the large band gap, wide depletion barrier, and doping limitations. In this work, we show that nanoscale heterostructure and polarization engineering of tunnel junctions can enable highly efficient tunneling even for ultra-wide bandgap materials where ordinary dopant-based homojunctions would have very low tunneling current density. In this letter, we discuss the design of AlGaN tunnel junctions over the entire Al composition range. We then experimentally demonstrate highly efficient tunnel junctions for AlGaN with Al composition greater than 50%.

Simulation of tunnel junctions was carried out using a two dimensional device simulator.[12] We assumed linear change of acceptor activation energy from 140 meV in GaN to 630 meV in AlN[13], constant donor activation energy of 15 meV for AlGaN with Al composition lower than 80%, and donor activation energy of 200 meV for Al compositions of 80% and greater[14]. Band diagrams were calculated by self-consistently solving Schrodinger-Poisson equations. Wentzel–Kramers–Brillouin (WKB) approximation was used for tunneling probability calculation, and tunneling current was calculated by considering all possible contributions by carriers with different energies. Bandtail states and bandgap narrowing effect due to heavy impurity doping are not included in the simulations, though they could lead to increase in the tunneling probability.[15]

To understand the design space for high composition AlGaN tunnel junctions, we investigated three tunnel junction structures. An $Al_{0.7}Ga_{0.3}N$ p-n tunnel junction with acceptor doping density of $N_A=5\times10^{19}$ $cm^{-3}$, and donor doping density of $N_D=1\times10^{20}$ $cm^{-3}$ was first simulated. Even with relatively high doping densities, the zero bias depletion width is 12 nm (energy band diagram in Figure 2(a)), which prevents effective interband tunneling.

To surmount the high tunneling barrier, we investigated the use of heterostructure and polarization engineering. A thin $In_yGa_{1-y}N$ layer is inserted between p- and n-type $Al_xGa_{1-x}N$ layers to reduce the tunneling barrier energy and width (Fig. 2(a)).[16-22] Due to the polarization discontinuity, sheet charges with density over $10^{13}$ $cm^{-2}$ are induced at the AlGaN/InGaN heterointerfaces and this reduces the tunneling barrier to less than 3 nm (Fig. 2(b)), leading to much higher tunneling probability compared to the homojunction tunnel junction.[9,18] However, wide depletion barriers are present in both n- and p-AlGaN sides due to the large band offsets between AlGaN and InGaN. Moreover, the high acceptor activation energy in AlGaN pushes the Fermi level into the forbidden gap even with heavy Mg doping, making it hard to align the valence band edge in the p-side to the conduction band edge in the n-side of the tunnel junction structure. This makes it challenging to achieve effective interband tunneling at low reverse bias. It is also found that the depletion width in the p-AlGaN layer increases under reverse bias (shown for 20 A/cm$^2$) leading to extra voltage drop to achieve effective interband tunneling.

To reduce the depletion-related tunneling barrier, we introduce compositionally graded AlGaN layers.[23-25] In the scheme investigated here, for each $Al_xGa_{1-x}N$ layers( Fig. 2(c)), the region adjacent to the tunnel junction is graded from $Al_xGa_{1-x}N$ to $Al_{x-0.1}Ga_{1-x+0.1}N$ on the p-side, and from $Al_{x-0.1}Ga_{1-x+0.1}N$ to $Al_xGa_{1-x}N$ on the n-side (Fig. 2(c)). The compositional grading leads to a gradient in the polarization (*P*), which creates a three dimensional (3D) polarization charge in the range of 2 ~ 3 ×10$^{19}$ cm$^{-3}$ ($\rho_{3D} = -\nabla \cdot P = -|P_{AlxGaN} - P_{Alx-0.1GaN}|/t$, where *t* is the grading distance).[23-25] The negative 3D polarization bound charge attracts high density of free holes, pushing the valence band close to the Fermi level in the graded p-AlGaN region. This results in a reduced depletion barrier in p-AlGaN (compared to the non-graded p-AlGaN), as confirmed by the energy band diagram in Fig. 2(c), leading to an increase in the tunneling probability even at small reverse bias.

As shown from the calculated current-voltage curves for $Al_{0.7}Ga_{0.3}N$ tunnel junctions in Fig. 3, the resistance for homojunction tunnel junction was found to be high, with 4 A/cm$^2$ under 50 V reverse bias, suggesting that unlike GaN[26-28], homojunctions may not be a viable solution for high composition ultra-wide bandgap AlGaN. In comparison, high tunneling current can be achieved for ultra-wide bandgap AlGaN using polarization engineered tunnel junction structure. More significantly, the graded tunnel junction structure leads to a further reduction in the reverse bias from 1.8 V to 0.12 V, and in the tunneling resistance from 2.8 ×10$^{-4}$ Ω cm$^2$ to 3.7 ×10$^{-5}$ Ω cm$^2$ at 1 kA/cm$^2$, both of which have important consequences for device efficiency.

To further explore the design space for ultra-wide bandgap tunnel junctions, various combinations of AlGaN/ InGaN compositions (constant thicknesses and doping levels same as that in Fig. 2(b) and 2(c)) were modeled. Fig. 4 shows the calculated tunneling resistance and reverse bias at 1 kA/cm$^2$ for tunnel junctions with and without compositional grading layers. InGaN interband tunneling barrier is dominant at low In compositions (< 10%), leading to high resistance for both graded and non-graded tunnel junctions. Lower tunneling resistance is observed when 10% to 80% In mole fraction is used, but the graded tunnel junction structure is critical to ensure lower tunneling resistance and voltage drop, especially for high Al composition (> 50%). In the structures that are feasible from lattice mismatch considerations (up to 30% In mole fraction), for the entire Al composition range, tunneling resistance in the ~ $10^{-4}$ $\Omega$ $cm^2$ range are achievable, with a voltage drop less than 1 V at 1 $kA/cm^2$. Moreover, the light absorption loss due to the ultra-thin InGaN layer is believed to be low.[9] Finite-difference time-domain (FDTD) calculations suggest an increase in the light extraction efficiency from ~ 14% to 60% (transverse-electric polarized light) and from ~ 8% to 40% (transverse-magnetic polarized light) when the p-GaN thickness is reduced from 20 nm to 4 nm in a conventional UV LED device with surface roughening.[29] Therefore, similar enhancement in light extraction efficiency is expected by replacing p-GaN with a tunnel junction given that the absorption coefficients of $In_yGa_{1-y}N$ ($0 \leq y \leq 0.3$) and GaN are similar for the emitted UV light.[30]

The tunnel junction designs were tested experimentally by integrating a graded tunnel junction on a UV LED structure as shown in Fig. 5. The structure was grown by $N_2$ plasma assisted molecular beam epitaxy (PAMBE) on an unintentionally-doped metal-polar $Al_{0.78}Ga_{0.22}N$ template with a substrate dislocation density of $7\times10^9$ $cm^{-2}$. The device structure consisted of a 600 nm $Al_{0.55}Ga_{0.45}N$ bottom contact layer with Si doping density of $1.2\times10^{19}$ $cm^{-3}$, three periods of 2 nm $Al_{0.4}Ga_{0.6}N$/ 6nm $Al_{0.55}Ga_{0.45}N$ quantum wells (QWs)/ barriers, 8 nm $Al_{0.72}Ga_{0.28}N$ p-type electron blocking layer, 25 nm p-$Al_{0.55}Ga_{0.45}N$, graded tunnel junction layer, and 300 nm n-type top contact layer. The graded tunnel junction structure consisted of 4 nm $In_{0.2}Ga_{0.8}N$, and AlGaN grading from 45% (55%) to 55% (45%) Al mole fraction on the n- (p-) side. Since the compositional grading was achieved by varying Al cell temperature, we estimate a grading thickness of approximately 10 nm. Self-consistent energy band calculations show that the p-AlGaN side is degenerate due to the polarization induced 3D charge.[24]

Light emitting diodes for on-wafer measurements were fabricated on the epitaxial structure described above.[9,11] Partial metal coverage (~ 35% electrode coverage for 30 × 30 μm$^2$ devices) for top contacts was used to minimize light absorption since light was collected from the top surface.[9,11] On-wafer electroluminescence (EL) and power measurements were carried out at room temperature under continuous wave mode.[11]

The measured current-voltage characteristic for the 30 × 30 μm$^2$ device is shown in Fig. 6. The device showed a current density of 10 A/cm$^2$ at 6.8 V. The additional voltage drop (2.4 V) over the quantum well band gap ($Al_{0.4}Ga_{0.6}N$ – 4.4 eV) is attributed to voltage drop across the electron blocking layer and the graded tunnel junction layer. The device reaches 1 kA/cm$^2$ at 11.2 V, with a differential resistance of $1.5\times10^{-3}$ Ω cm$^2$, which is one order lower compared to the on-resistance of reported conventional UV LED devices (~ $2\times10^{-2}$ Ω cm$^2$ ) with similar emission wavelengths.[31-34] The top contact resistance is extracted to be $5.6\times10^{-4}$ Ω cm$^2$ based on transfer length method measurements. The bottom contact resistance and spreading resistances are neglected. By subtracting the contact resistance from the total series resistance, it gives an upper bound of the tunnel junction resistance of $9.4\times10^{-4}$ Ω cm$^2$ at 1 kA/cm$^2$.

The experimental (red stars) tunneling resistance and voltage drop of the $Al_{0.55}Ga_{0.45}N$/ $In_{0.2}Ga_{0.8}N$ tunnel junctions and previous graded $Al_{0.3}Ga_{0.7}N$/ $In_{0.25}Ga_{0.75}N$[9], GaN/ $In_{0.25}Ga_{0.75}N$[18], and GaN p+-n+[28] tunnel junctions are shown together with the simulation predictions in Fig. 4. The experimental values are of the same order of magnitude as the simulated results for the tunnel junction structures. The discrepancy could be attributed to the approximations made in the simulations, as well as non-idealities in growth and material quality.

Tunneling hole injection was confirmed by light emission from the devices. As shown in Fig. 7(a), single peak emission at 292 nm was achieved across the entire device area, with emission power of 0.1 mW at 6 mA injection current for the 30 × 30 μm$^2$ device (Fig. 7(a) inset). The maximum external quantum efficiency (EQE) and wall-plug efficiency values were 0.41% and 0.16%, respectively. The output power is underestimated since no integrating sphere was used in the electroluminescence measurement and only the light emitting from the top surface was collected.[11] However, the external quantum efficiency and wall-plug efficiency did not show droop until high current density, suggesting high non-radiative recombination in the active region associated with native defects or dislocations.[35] The device efficiencies achieved are comparable to state-of-the-art results (EQE = 0.18% ~ 1%)[31-34] at similar wavelengths even though the output power is underestimated due to on-wafer measurement.

In conclusion, we have demonstrated the design of graded tunnel junction structure to achieve low tunneling resistance with low voltage drop. Polarization induced 3D charge is beneficial in reducing tunneling barrier especially for high Al content AlGaN tunnel junctions. Non-equilibrium hole injection into wide bandgap materials with bandgap larger than 4.7 eV was demonstrated using a tunneling injected UV LED structure, and light emission at 292 nm was achieved based on tunneling injection. The work described here demonstrates the feasibility of using tunneling for hole injection into ultra-wide bandgap materials, and could enable a new class of highly efficient optical and electronic devices based on the III-Nitride system.

Acknowledgement: We acknowledge funding from the National Science Foundation (ECCS-1408416). Sandia National Laboratories is a multi-program laboratory managed and operated by Sandia Corporation, a wholly owned subsidiary of Lockheed Martin Corporation, for the U.S. Department of Energy's National Nuclear Security Administration under contract DE-AC04-94AL85000.

Figure captions:

Fig. 1 Schematic structures of (a) a typical conventional UV LED and (b) a tunneling injected UV LED.

Fig. 2 Comparison of tunnel junction structures formed by (a) heavily doped p-n homojunction, (b) heterojunction with thin InGaN layer inserted between AlGaN layers, (c) heterojunction further with graded Al composition layers. The Al/ In mole fraction profiles, schematic charge profiles and band diagrams of $Al_{0.7}Ga_{0.3}N$ (/ $In_{0.3}Ga_{0.7}N$) tunnel junctions are compared.

Fig. 3 Calculated current-voltage curves for the $Al_{0.7}Ga_{0.3}N$ (/ $In_{0.3}Ga_{0.7}N$) tunnel junctions (correspond to the band diagrams in Fig. 2). The homojunction tunnel junction is highly resistive, while polarization engineered tunnel junction structures enable much lower tunneling resistances.

Fig. 4 Simulated tunneling resistance and reverse voltage drop at a current density of 1 kA/cm$^2$ for AlGaN/ InGaN tunnel junction structures without and with the compositional grading layers. Experimental results of graded $Al_{0.55}Ga_{0.45}N$/ $In_{0.2}Ga_{0.8}N$, graded $Al_{0.3}Ga_{0.7}N$/ $In_{0.25}Ga_{0.75}N$[9], GaN/ $In_{0.25}Ga_{0.75}N$[18], and GaN p+-n+[28] tunnel junctions are marked in the figures.

Fig. 5 Epitaxial stack, and equilibrium energy band diagram of tunneling injected $Al_{0.55}Ga_{0.45}N$ UV LED structure.

Fig. 6 Electrical characteristics of the tunneling injected $Al_{0.55}Ga_{0.45}N$ UV LED devices (30 × 30 μm$^2$).

Fig. 7 (a) Electroluminescence of the tunneling injected $Al_{0.55}Ga_{0.45}N$ UV LED structure with dc current injection from 0.3 mA to 6 mA at room temperature, showing single peak emission at 292 nm. (b) EQE and WPE of the 30 × 30 μm$^2$ device. The inset to (a) shows the output power, and the inset to (b) is an optical micrograph image of a device (30 × 30 μm$^2$) operated at 5 mA.

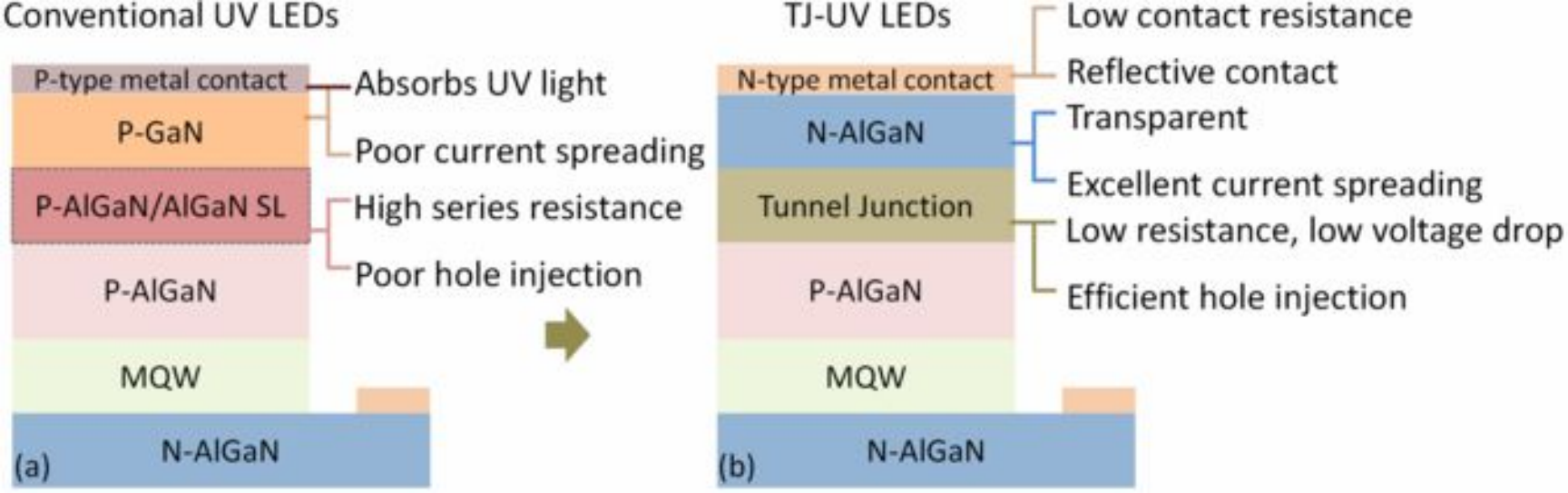

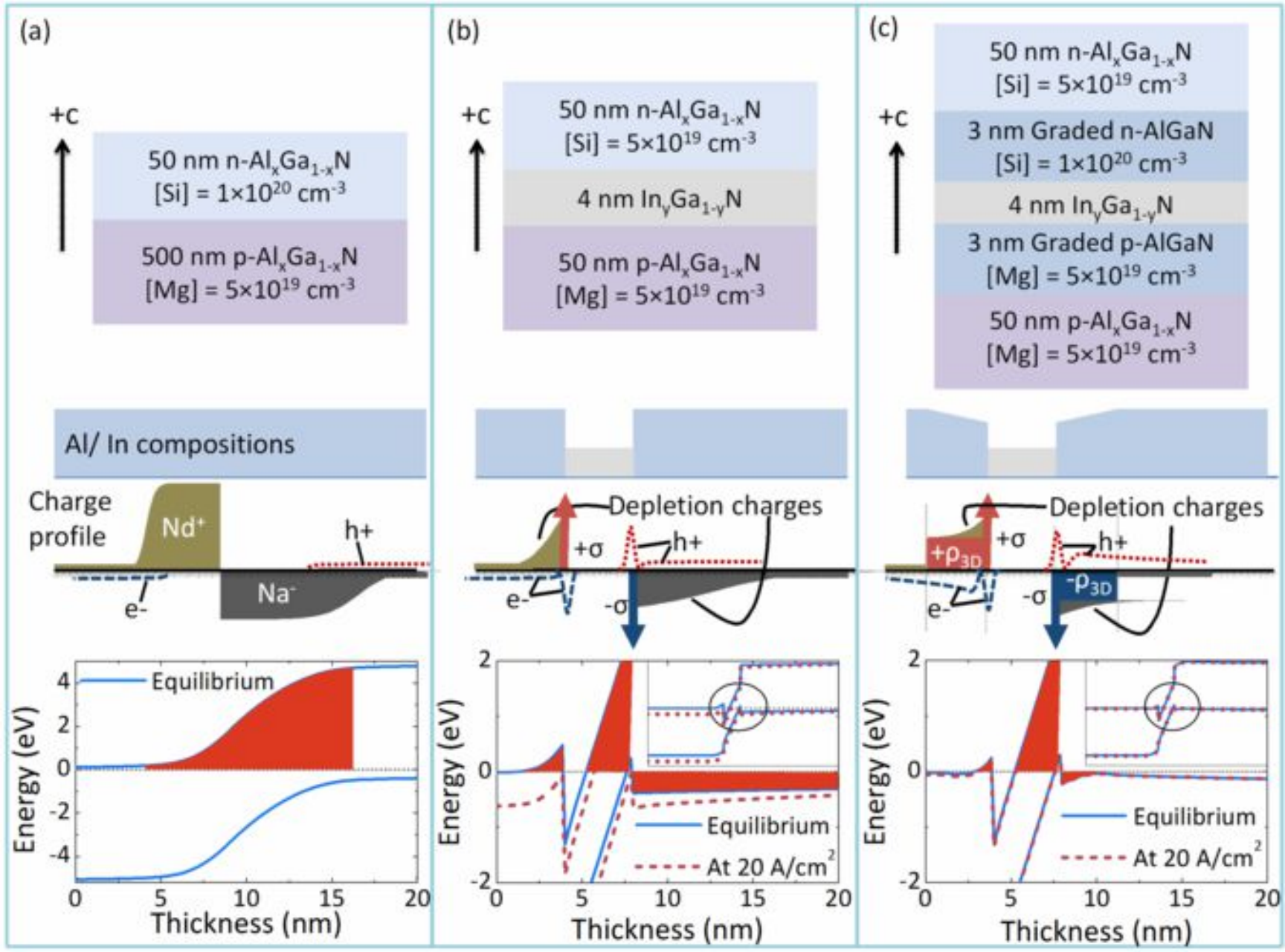

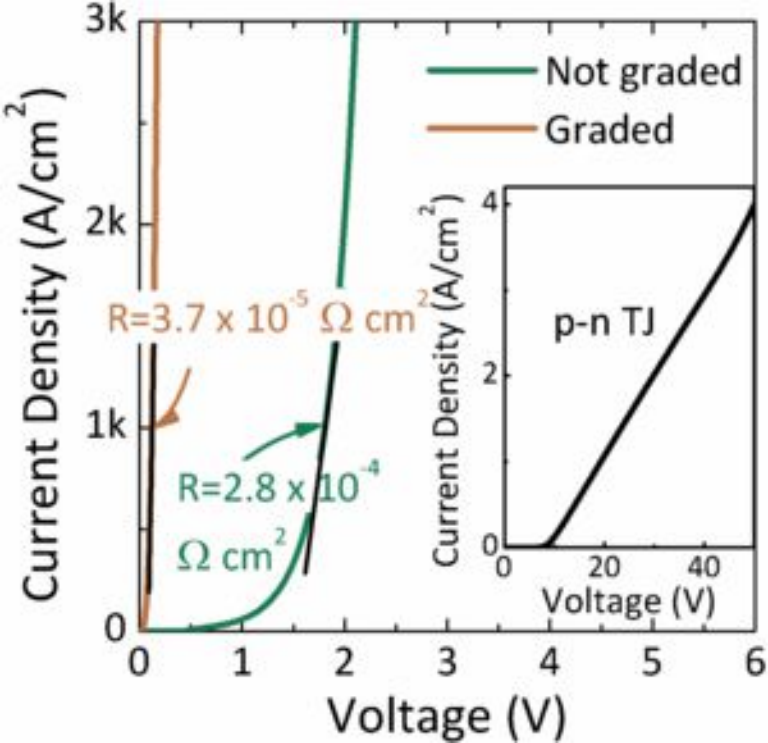

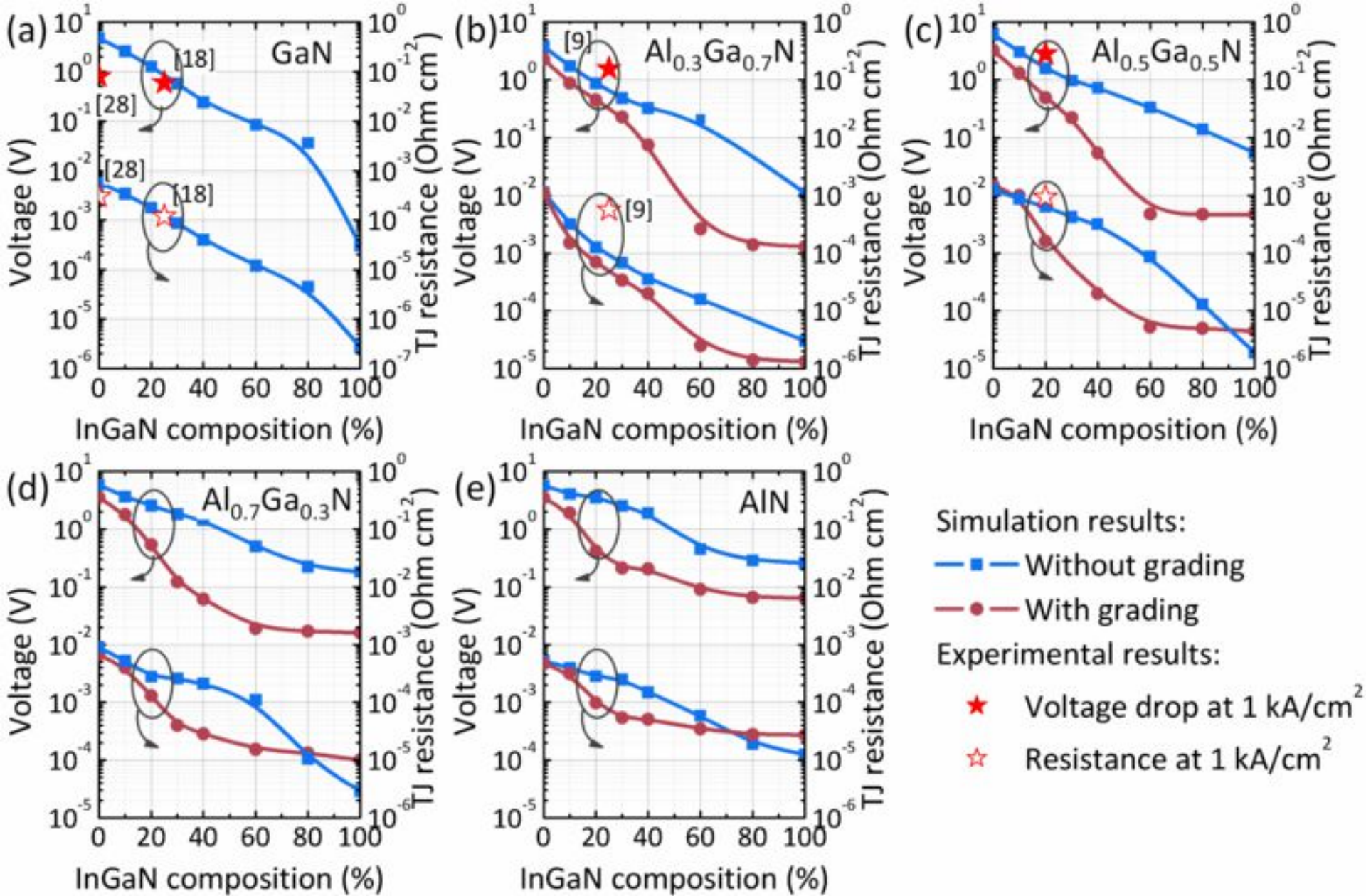

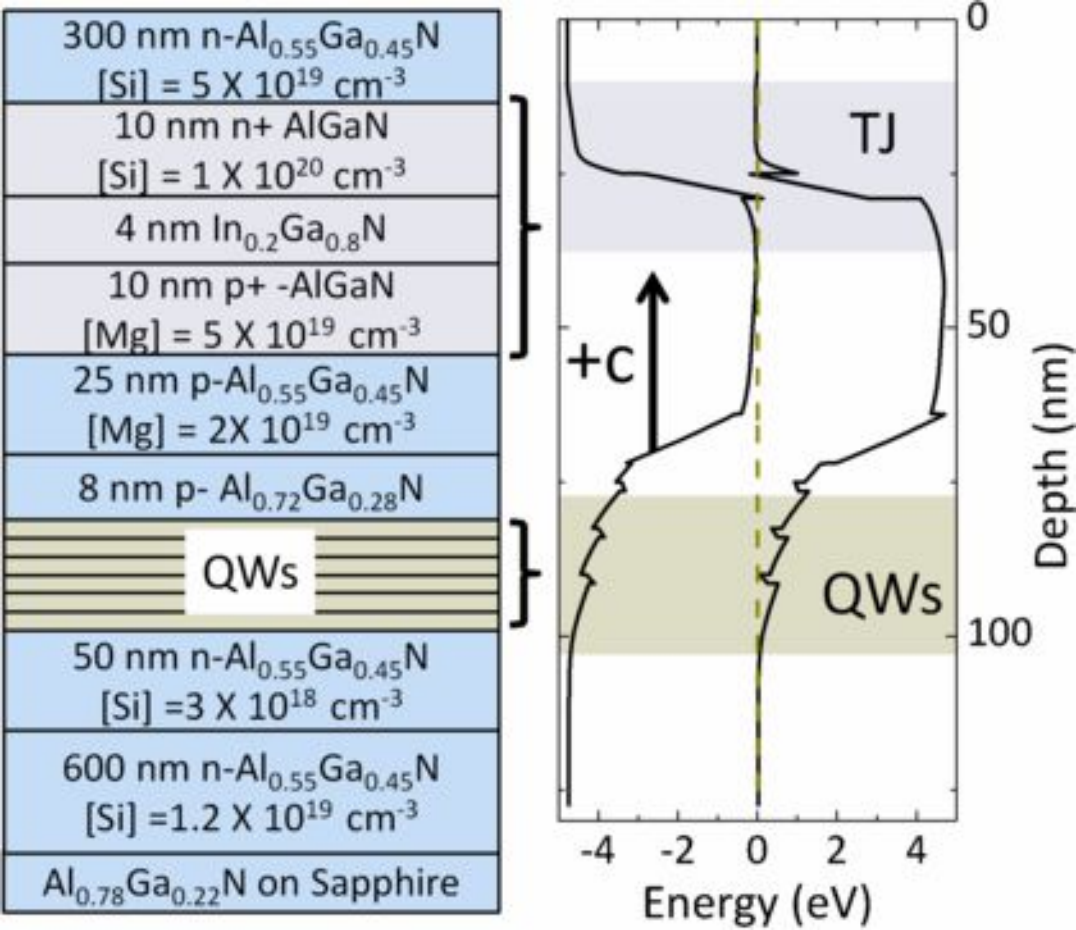
300 nm n-$Al_{0.55}Ga_{0.45}N$
[Si] = 5 X $10^{19}$ $cm^{-3}$
10 nm n+ AlGaN
[Si] = 1 X $10^{20}$ $cm^{-3}$
4 nm $In_{0.2}Ga_{0.8}N$
10 nm p+ -AlGaN
[Mg] = 5 X $10^{19}$ $cm^{-3}$
25 nm p-$Al_{0.55}Ga_{0.45}N$
[Mg] = 2X $10^{19}$ $cm^{-3}$
8 nm p- $Al_{0.72}Ga_{0.28}N$
QWs
50 nm n-$Al_{0.55}Ga_{0.45}N$
[Si] =3 X $10^{18}$ $cm^{-3}$
600 nm n-$Al_{0.55}Ga_{0.45}N$
[Si] =1.2 X $10^{19}$ $cm^{-3}$
$Al_{0.78}Ga_{0.22}N$ on Sapphire
TJ
+C
QWs
0
50
100
Depth (nm)
-4
-2
0
2
4
Energy (eV)

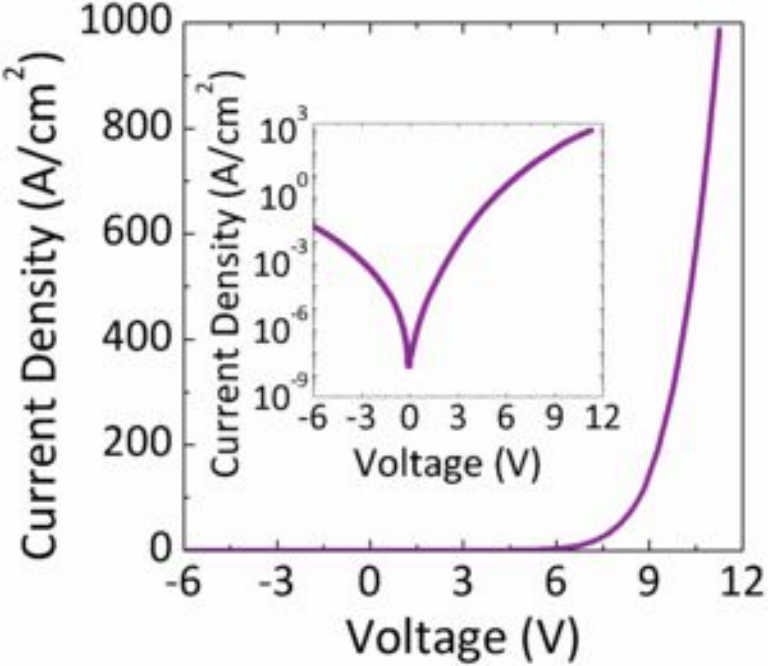

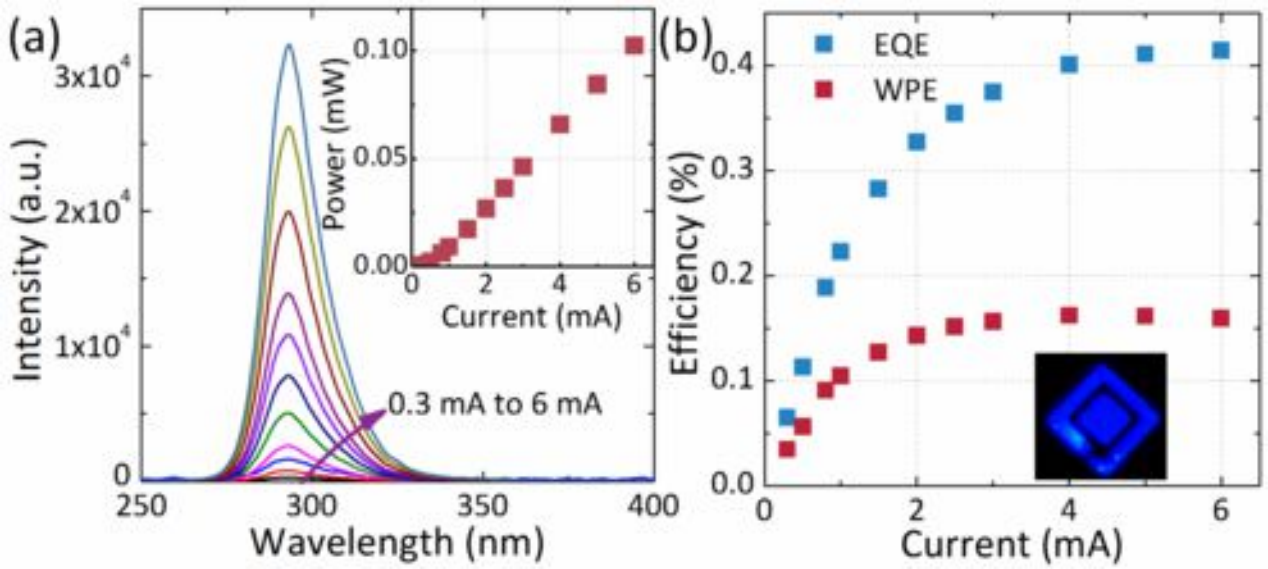